\documentclass[useAMS,usenatbib]{mn2e}

\usepackage {graphicx}
\usepackage {times}
\usepackage{pdflscape}
\usepackage{rotating}

\newcommand{\lya}{Ly\,$\alpha$~}
\newcommand{\lyb}{Ly\,$\beta$~}
\newcommand{\lyg}{Ly\,$\gamma$~}
\newcommand{\lyd}{Ly\,$\delta$~}

\title[A Gunn-Peterson test with a QSO at z=6.4]{A Gunn-Peterson test with a QSO at z=6.4\footnotemark[0]
}
\author[Goto]{Tomotsugu Goto$^{1,2}$ 
 \thanks{E-mail:tomo@ifa.hawaii.edu} \thanks{JSPS SPD Fellow},
 Yousuke Utsumi$^{3}$,
Takashi Hattori$^{1}$,
\newauthor 
Satoshi Miyazaki$^{3}$,
and Chisato Yamauchi$^{4}$
\\
$^{1}$Subaru Telescope 650 North A'ohoku Place Hilo, HI 96720, USA\\
$^{2}$Institute for Astronomy, University of Hawaii
2680 Woodlawn Drive, Honolulu, HI, 96822, USA\\
$^{3}$ Department of Astronomical Science,The Graduate University for Advanced Studies,
 2-21-1 Osawa, Mitaka, Tokyo 181-8588, Japan\\
$^{4}$Institute of Space and Astronautical Science, Japan Aerospace Exploration Agency, 	     Sagamihara, Kanagawa 252-5210
}

\begin{document}
\def\Hg{H$\gamma$}
\def\Hd{H$\delta$}

\date{\today; in original form 2011 Feb 7}

\pagerange{\pageref{firstpage}--\pageref{lastpage}} \pubyear{2009}

\maketitle

\label{firstpage}

\begin{abstract}
 Understanding the cosmic re-ionization is one of the key goals of the modern observational cosmology.
 High redshift QSO spectra can be used as background light sources to measure absorption by intervening neutral hydrogen. 
 We investigate neutral hydrogen absorption in a deep, moderate-resolution Keck/Deimos spectrum of QSO CFHQSJ2329-0301 at z=6.4.
 This QSO is one of the highest redshift QSOs presently known at z=6.4 but is 2.5 mag fainter than a previously well-studied QSO SDSSJ1148+5251 at z=6.4.
 Therefore, it has a smaller Stromgren sphere, and  allows us to probe the highest redshift hydrogen absorption to date.
 The average transmitted flux at $5.915<z_{abs}<6.365$ (200 comoving Mpc) is consistent with zero, in Ly $\alpha$, Ly $\beta$, and \lyg absorption measurements. 
 This corresponds to the lower limit of optical depth, $\tau_{eff}^{\alpha}>4.9$.
 These results are consistent with strong evolution of the optical depth at $z>5.7$.
\end{abstract}


\section{Introduction}
 Understanding when and how the neutral dark Universe was re-ionized by the first stars, galaxies, and black holes is a key goal of modern observational cosmology. The bright UV continuum of high-redshift QSOs can be used as a background light source to probe this cosmic re-inonization since it is absorbed by neutral hydrogen. 
 Previously, \citet[][F06 hereafter]{2006AJ....132..117F} compiled 19 QSOs at $5.74\leq z<6.42$ to find the sudden increase of optical depth ($\tau$) and  the length of dark gaps at z$\sim$6, suggesting an approach to the end of the re-ionization. 

However,  increasing diversity has been observed in the optical depth in various line of sights, suggesting it is important to use a statistically significant sample size to discuss the evolution of optical depth. The highest redshift QSO previously used for absorption analysis, SDSSJ1148+5251 at z=6.42, had escaping flux detected at z=6.25 \citep{2003AJ....125.1649F}. 
This means there is still room for the QSO absorption measurement to be a useful probe of the cosmic re-ionization.

In this work, we observe  QSO CFHQSJ2329-0301 at z=6.4 with the Keck/Deimos spectrograph.
This QSO is $\approx$2.5 mag fainter than the previously well-studied highest redshift QSO SDSSJ1148+5251 at z=6.4.
Therefore, the QSO has a few times smaller Stromgren sphere (in radius), allowing us to measure optical depth much closer to the QSO emission at higher redshift.
We measure the neutral hydrogen absorption in this QSO spectra to gain a better understanding of the evolution of the optical depth at high redshift.

  Unless otherwise stated, we adopt the WMAP cosmology: $(h,\Omega_m,\Omega_L) = (0.7,0.3,0.7)$ \citep{2011ApJS..192...18K}.

%

\begin{figure*}
\begin{center}
\includegraphics[scale=0.9]{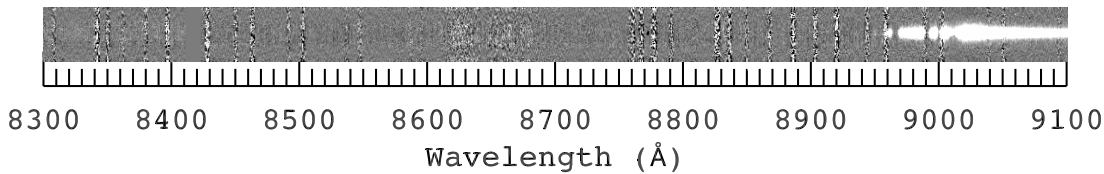}
\end{center}
\caption{
Two dimensional spectra of CFHQSJ2329-0301, taken with the Keck/Deimos spectrograph. One-dimensional spectrum is shown in Fig. \ref{fig:1d}.
}\label{fig:2d}
\end{figure*}

%

\begin{figure*}
\begin{center}
\includegraphics[scale=0.9]{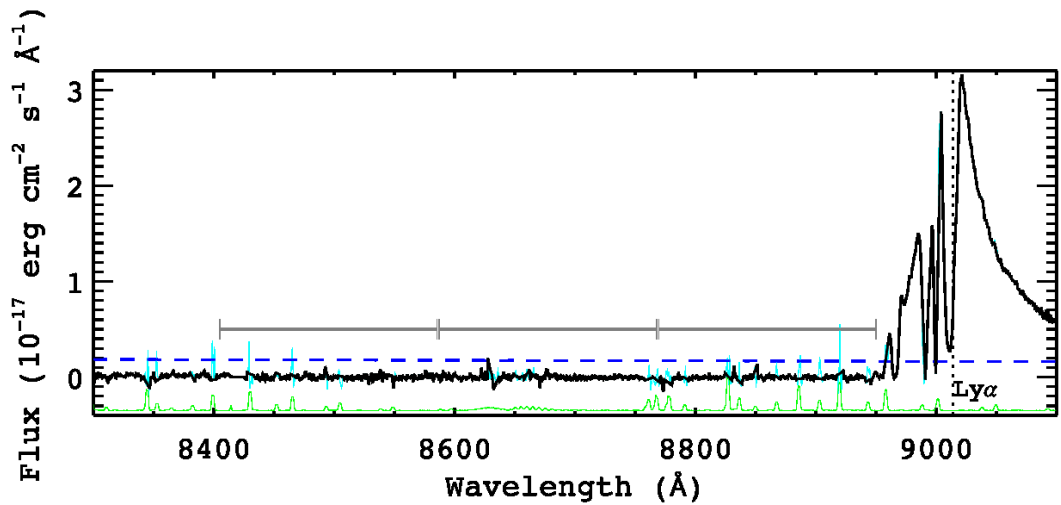}
\end{center}
\caption{
One-dimensional spectra of CFHQSJ2329-0301. 
 The blue dashed line shows the best-fit continuum level, assuming a power law of $f_{\nu}\propto$ $\nu ^{-0.5}$.
The cyan lines show the wavelengths masked out due to the strong sky emissions.
The green lines show the sky spectrum in arbitrary units.
The gray lines show the first three wavelength regions where the flux was consistent with zero.
The corresponding two-dimensional spectrum is shown in Fig. \ref{fig:2d}.
}\label{fig:1d}
\end{figure*}



\section{Observation}
\label{Observation}

%
%

Our target is a QSO CFHQSJ2329-0301\citep[][W07 hereafter]{2007AJ....134.2435W}. 
Initially the redshift was reported to be z=6.43 based on Ly $\alpha$ (W07). 
The redshift was later corrected to be z=6.417 based on the Mg II line  \citep{2010AJ....140..546W} because blueward of the Ly $\alpha$ emission line suffers from heavy absorption from neutral hydrogen. 
 This is one of the three highest redshift QSOs to date, along with SDSSJ1148+5251 at z=6.419 \citep{2003AJ....125.1649F}  and CFHQS0210-0456 at z=6.438 \citep{2010AJ....140..546W}. 
This QSO is known to be in a dense environment surrounded by 7 Lyman break galaxies (LBGs) \citep{2010ApJ...721.1680U}  and to have extended Ly $\alpha$ emission and continuum \citep{2009MNRAS.400..843G}.
Characteristics of the target are summarized in Table \ref{tab:targets}. 

We obtained moderate resolution spectra of the QSO using Keck/Deimos \citep{2003SPIE.4841.1657F}  on Sep. 12-13, 2010.
 Although Deimos was used in the multi-slit mode to investigate the environment of the QSO, we focus on the QSO spectra in this paper.
 We used the 830 lines mm$^{-1}$ grating and the OG550 filter with a central wavelength of 8000\AA~. This setting has a wavelength coverage of 6000\AA\ to 10000\AA\ with a spectral resolution of $\sim$2.5\AA~ for 0.75'' slit. The spatial resolution is 0.1185'' pixel$^{-1}$. 
The total integration time was 5.5 hours. 
 
 We used the DEEP2 pipeline \citep{2004AJ....127.3121W,2007ApJ...660L...1D} to reduce the data, except the background subtraction, which we did manually to remove ghost features of the 830 grating.
 Wavelength calibration is based on the HeNeAr lamp. 
The spectrophotometric standard G191-B2B was observed and used for flux calibration.
Fig. \ref{fig:2d} shows two-dimensional spectrum of the QSO. 

 
At $>$7500\AA, many strong sky OH emission lines are present. Although the DEEP2 pipeline subtract these sky lines well, the noise level at the exact wavelengths of the sky lines are significantly higher than the other wavelength clear of emission. Therefore, we have masked out these wavelengths of strong sky emissions from the absorption measurement. In Fig. \ref{fig:1d},
the wavelengths masked out are shown in thin cyan lines. The green lines at the bottom show the sky spectrum in arbitrary units. We checked that this procedure only affects our measurements at the few per cent level, and does not change the main conclusion of the paper.


We used double Gaussians to fit  Ly $\alpha$ and NV lines. 
The double-gaussian model do not fit the observed spectra well due to the heavy absorption blueward of Ly$\alpha$ and broad natures of both lines. The best-fit rest-frame equivalent widths are 
110$\pm$60\AA, and 6$\pm$1\AA~ for Ly $\alpha$ and NV, respectively. 
These values are consistent with previously reported values (W07).

The spectrum shows four strong Ly $\alpha$ absorptions on the blue-side of the emission (Fig. \ref{fig:1d}). 
There also exist absorption lines at 
9183,
9212,
and 9240\AA~
with equivalent widths of
1.4, 1.5 and 2.6\AA, respectively.
Because none of the pairs corresponds to different lines of the same system (such as  SiII$\lambda$1260, SiII$\lambda$1304, OI$\lambda$1302, and CII$\lambda$1334), the identities of these lines are not clear.
There also is an emission line at 9681\AA, corresponding to OI line (1302.17\AA) at z=6.43.

%

\begin{figure}
\begin{center}
\includegraphics[scale=0.6]{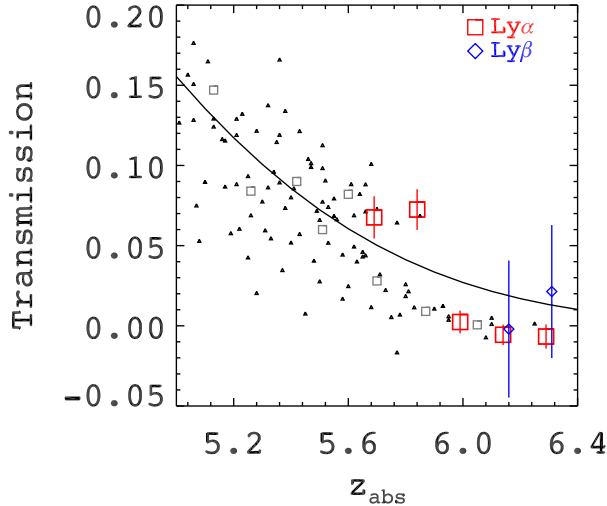}
\end{center}
\caption{
Lyman transmission in the spectrum of CFHQSJ2329-0301 (red, filled-circles).
The black triangles are from \citet{2006AJ....132..117F}, and the gray squares are from \citet{2004AJ....127.2598S}.
}\label{fig:trans}
\end{figure}

\section{Absorption measurements and results}\label{results}

High redshift QSOs can be used as a background light source to investigate the inter-galactic medium (IGM) at high redshift.
In the following subsections, we measure average absorption in the Ly$\alpha$, \lyb, and \lyg forest regions.

\begin{table*}
 \begin{minipage}{180mm}
  \caption{Target information adopted from \citet{2007AJ....134.2435W} and \citet{2009MNRAS.400..843G}. }\label{tab:targets}
  \begin{tabular}{@{}lcccccccc@{}}
  \hline
   Object &  z$_{\rm Mg II}$&              $i'_{AB}$ & $z'_{AB}$ & $z_{R}$ & $J$                                  &  \\ 
 \hline
 \hline
CFHQS J232908.28-030158.8  & 6.417$\pm$0.002 & 25.54$\pm$0.02 & 21.165$\pm$0.003&  21.683$\pm$0.007          & 21.56$\pm$0.25 \\
\hline
\end{tabular}
\end{minipage}
\end{table*}

To measure the IGM absorption, we need to determine the continuum before the absorption.
This is not a trivial process because the continuum blueward of the Ly$\alpha$ is heavily absorbed by the neutral hydrogen. 
At the same time, our optical CCD loses sensitivity at around 10,000\AA, and we do not have a long baseline to fit continuum, let alone the presence of strong Ly$\alpha$, NV and OI emissions.
Following previous work by \citet{2000AJ....119..928F,2001AJ....122.2833F}, we assume a power-law continuum of $f_{\nu} \propto \nu^{-0.5}$. 
When the continuum slope is uncertain, the slope of $\nu^{-0.5}$ is commonly used by previous works. This is the case for most QSOs at z$\sim$6.
 We fit the continuum of $\nu^{-0.5}$ to the spectrum at $>9300$\AA~to determine the continuum level. 
 We also follow  \citet{2000AJ....119..928F,2001AJ....122.2833F} to consider variation from  $\nu^{+0.5}$ to $\nu^{-1.5}$ to estimate errors of $\tau$.
 However, this is a conservative approach since at least at lower z, most QSO continuum slopes are between  -0.3 and -0.9 \citep[e.g.,][]{2001AJ....122..549V}. Some recent papers do not even include variation of the slope in the errors of $\tau$.
 In this work, errors in the continuum fit and the variation in the slope are included in the final uncertainty in the transmission.
Based on this continuum, we measure average transmission in the following narrow wavelength (redshift) windows.
Then, the transmission is used to calculate the effective optical depth.


\subsection{\lya absorption}

With QSO CFHQSJ2329-0301, we can measure absorption closer to the Ly$\alpha$ emission than bright QSOs, 
because this QSO has a small Stromgren sphere (ionized zone) of 6.4 Mpc (where Ly$\alpha$ flux 
drops first to $T<0.1$ for spectra binned in 20\AA~ pixel$^{-1}$) due to its faint intrinsic luminosity (W07).
To measure absorption, we used the maximum redshift, $z_{max}$, of 6.365, where blueward of continuum does not show any sign of rise.
 This gives us  $z_{em}- z_{max}=0.052$.
 This is the highest redshift absorption measurement from the QSO spectra to date due to the high redshift of QSO (z=6.4) and its small Stromgren sphere.

We measure Ly$\alpha$ absorption down to 1040\AA, which is the shortest wavelength that is not affected by the Ly$\beta$+OVI emissions.
Within this range, we used the redshift interval of $\Delta z$ of 0.15, which corresponds to $\sim$60 Mpc in comoving distance, for comparison with previous work.
Although the resolution of our spectrum is lower than high-resolution spectra used in the past, this is not a disadvantage for a binned measurement like the transmission.
The first three bins are shown in gray horizontal lines in Fig. \ref{fig:1d}.
The median redshifts of five bins are presented in Table \ref{tab:transtau}.
 Using this redshift range and interval, we measure the transmission as 

\begin{math}
\mathcal{T}(z_{abs})\equiv <f_{\nu}^{obs}/f_{\nu}^{int}>\label{trans}
\end{math}

The measured transmitted flux ($\mathcal{T}$) is shown in Table \ref{tab:transtau} and plotted in Fig. \ref{fig:trans}.
 We estimate errors by changing the continuum slope from $\nu^{+0.5}$ to $\nu^{-1.5}$, quadruply added by the noise in spectra and the errors in determining the continuum level.  The errors on $\mathcal{T}$ are dominated by the continuum determination.
 
 In Fig. \ref{fig:trans} and Table \ref{tab:transtau}, the first three bins at $5.915<z<6.365$ have no detected flux and are consistent with zero within the error.
 Only two lower redshift bins at $5.615<z<5.915$ have detected flux.

\begin{figure}
\begin{center}
\includegraphics[scale=0.6]{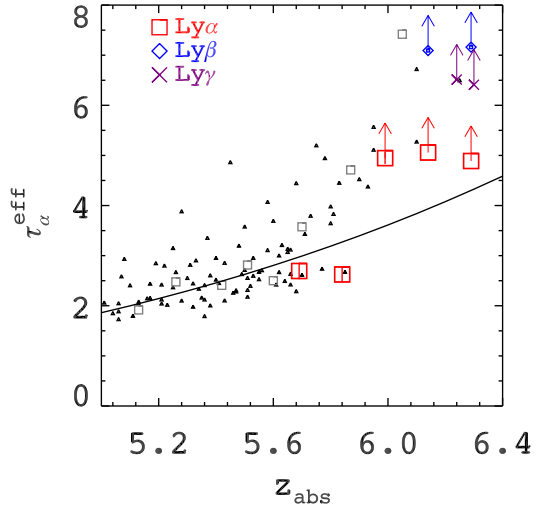}
\end{center}
\caption{
Effective Gunn-Peterson optical depth from the spectrum of CFHQSJ2329-0301 (red, filled-circles). 
The black triangles are from \citet{2006AJ....132..117F}, and the gray squares are from \citet{2004AJ....127.2598S}.
The solid line is the best power-law fit to the data at $z<5.5$ by F06 (eq.5).
}\label{fig:tau}
\end{figure}

 It is conventional to express optical depth $\tau$ in terms of $\tau=-ln (\mathcal{T})$.
 We present $\tau$ as a function of redshift in Fig. \ref{fig:tau}. 
 For bins where we do not have detected flux, we convert the errors on $\mathcal{T}$ to the lower limit of the  $\tau$. 

In Figs.\ref{fig:trans} and \ref{fig:tau}, small triangles and gray squares represent previous data from F06 and \citet{2004AJ....127.2598S}.
The solid line shows the best power-law fit to the data at $z<5.5$ by F06 (eq.5).



 
\begin{table}
\begin{center}
\caption{IGM absorption  towards the QSO CFHQSJ2309-0301 at $z=6.42$.  The widths of bins are  $\Delta z=0.15,0.15$, and 0.06 for \lya, \lyb, and \lyg~  absorptions, respectively.}
\label{tab:transtau}
\begin{tabular}{ccrl}
  \hline
   Redshift & Line & Transmission & $\tau_{\alpha}$  \\ 
 \hline
6.29  & \lya &-0.007$\pm$0.008 &    $>$4.9\\
6.14  & \lya &-0.006$\pm$0.006 &    $>$5.1\\
5.99  & \lya & 0.002$\pm$0.007 &    $>$4.9\\
5.84  & \lya & 0.073$\pm$0.013 &    2.6$\pm$0.1\\
5.69  & \lya & 0.068$\pm$0.013 &    2.7$\pm$0.2\\
\hline
6.29  & \lyb &  0.021$\pm$0.041 &    $>$7.2    \\
6.14  & \lyb & -0.002$\pm$0.043 &    $>$7.1    \\
\hline
6.29  & \lyg &   0.26$\pm$0.23 &    $>$6.4\\
6.23  & \lyg &  -0.08$\pm$0.23 &    $>$6.5\\
\hline
\end{tabular}
\end{center}
\end{table}

\subsection{\lyb absorption}

 At the same neutral hydrogen density, the optical depth $\tau$ is proportional to $f\lambda$, where $f$ and $\lambda$ are the oscillator strength and rest-frame wavelength of the transition. Therefore, the $\tau$ of Ly$\beta$ is factor of 6.2 smaller than that of Ly$\alpha$. Thus, Ly$\beta$ can probe into much more neutral hydrogen than Ly$\alpha$ absorption.
 In this section we measure optical depth using Ly$\beta$ absorption from the spectra. 

We assume the same continuum as was used for Ly$\alpha$. We choose the minimum redshift of Ly$\beta$ as 970\AA, above which it is not affected by Ly$\gamma$ absorption. This leaves us the redshift range of $6.065<z_{abs}<6.365$.
 The Ly$\beta$ absorptions overlap with Ly$\alpha$ absorption from lower redshift. Therefore, the \lya absorption has to be corrected in measuring $\tau^{\beta}$.
 We used eq. (5) of F06 to estimate Ly$\alpha$ absorption from lower redshift, and corrected the Ly$\beta$ transmission measurements.

 Table. \ref{tab:transtau} shows the results. They are also plotted in Fig. \ref{results} with the blue diamonds.
 No flux is detected in Ly$\beta$ measurements. The two measured \lyb transmissions at $6.065<z_{abs}<6.365$ are consistent with zero flux within the error.

 To convert Ly$\beta$ transmission to $\tau_{\beta}$, one has to consider different optical depths between \lya and \lyb.
 The difference depends on the UV background, clumpiness of the IGM, its equation of state, and the uniformity of the UV background.
 According to simulations \citep[e.g.,][]{2005ApJ...620L...9O} and empirical measurements (F06), the $\tau_{\alpha}/\tau_{\beta}$ conversion is in the range of 2.5-2.9.  The $\tau_{\alpha}/\tau_{\gamma}$ conversion lies in the range of 4.4-5.7.
Following the discussion in F06, we use  $\tau_{\alpha}/\tau_{\beta}$=2.25 and  $\tau_{\alpha}/\tau_{\gamma}$=4.4. 
 
 Fig. \ref{fig:tau} shows constraints on the $\tau_{\alpha}$ from \lyb absorption measurements. Note that $\tau$ is converted to \lya optical depth in Fig. \ref{fig:tau}. Since flux are not detected in both bins, Table \ref{tab:transtau} shows the lower limit in  $\tau_{\alpha}$ from the error.

\subsection{\lyg absorption}
 Similarly to \lyb, the \lyg optical depth is a factor of 17.9 smaller than that of \lya, providing us with a chance to probe into more neutral hydrogen.
 However, the \lyg absorption measurement is restricted by the presence of \lyd absorption at the low-redshift end and the QSO flux at the high-redshift end.
Therefore, we use a smaller binsize of $\Delta$z of 0.06. 
The \lyg transmission measurements are corrected for overlapping lower redshift \lya and \lyb absorptions using the best-fit power laws to lower redshift data (eqs. 5 and 6 in F06).
The optical depth is converted to the $\tau_{\alpha}$, using  $\tau_{\alpha}/\tau_{\gamma}$=4.4.
The results are in Table \ref{tab:transtau} and Fig. \ref{fig:tau}. 

The \lyg transmission has more potential since it has a factor of 17.9 smaller optical depth. 
 However, the \lyg absorption measurements are much more challenging.
 Since the overlapping foreground \lya and \lyb lines absorb $\sim$98\% of the continuum flux in the wavelength ranges of \lyg, 
 so we need to measure absorption in the remaining $\sim$2\%.
 Errors in continuum determination are also larger since we use the bluer part of the spectrum.
By just changing the slope of $\nu^{+0.5}$ to  $\nu^{-1.5}$, we get 10\% error in the measurement.
Combined with the noise in the spectra, we did not obtain tighter constraints than \lyb.
The purple crosses in Fig. \ref{fig:tau} show the lower limit from the measurement error.

\section{Discussion}\label{discussion}

 It has often been suggested that QSOs are an ineffective re-ionization probe because Ly$\alpha$ absorption saturate at neutral hydrogen fractions of $f_{HI} \sim10^{-4}$.
 However the Gunn-Peterson optical depths of Ly$\beta$ and Ly$\gamma$ are factors of 6.2 and 17.9 smaller than that of Ly$\alpha$. 
 It has also been suggested that metal lines such as OI can be used to probe into even larger optical depths \citep{2006ApJ...640...69B,Becker2011}.
 While the dark gap statistic method can also probe into more neutral hydrogen \citep{2002AJ....123.2183S,2002AJ....123.2151P}, our medium resolution spectrum did not allow us to measure the dark gaps.
Although we did not detect any flux at  $5.915<z_{abs}<6.365$, another z=6.4 QSO had a flux detected at z=6.29 with \lyg (F06).
 Therefore, there is room for high-z QSOs to be an important re-ionization probe beyond z=6.4. 
 Also in both Figs. \ref{fig:trans} and \ref{fig:tau},  $\tau$ and $\mathcal{T}$ show a significant variation, which suggests the importance of finding more bright QSOs at $z>6.4$ as an end of re-ionization probe. On-going and future imaging surveys at $>1\mu$m such as UKIDSS, VISTA, CFHQSIR, and PanSTARRS are expected to find such QSOs in the near future.

 One of the remaining problems is an uncertainty in continuum determination. 
 We used a slope of $\nu^{-0.5}$, but  $\nu^{+0.5}$ and $\nu^{-1.5}$ also fit continuum at $>$9300\AA~ well.
 Variation from these is larger than the errors on the flux, even with our relatively short exposure of a faint QSO.
 Ironically, it is hard to measure intrinsic continuum blueward of Ly$\alpha$ emission because of the heavy absorption that we aim to measure.
 To make the matter worse,  observations of a larger sample of QSOs at lower redshift show a continuum break at 1200$<\lambda<$1300\AA, with a steeper power-laws of  $\nu^{-1.7}$ at shorter wavelengths \citep{2002ApJ...565..773T}. If there is an evolution, the continuum blueward of high-z QSOs remain uncertain. Therefore, the continuum determination is a problem in measuring $\tau$ even if one takes very high S/N spectra. 

  Interpretation of the evolution of $\tau$ is a matter of active debate. 
 At $z>5.7$, F06 reported a departure of $\tau$ from the best-fit power law from $z<5.5$, arguing that $z\sim$6 is the end of the reionization, where the Universe was not completely ionized. The average length of dark absorption gap also jumps from $<$10 to $>$80 comoving Mpc at $z>5.7$. 
Our $\tau$ measurements are clearly above the solid line (fit to the $z<5.5$ data) in Fig. \ref{fig:tau}, and thus are consistent with this sudden increase of $\tau$ at $z>5.7$.
On the other hand, \citet{2007ApJ...662...72B} showed that the probability distribution of \lya transmitted flux of QSOs at $2<z<5$ are better fit by lognormal optical depth distributions. This lognormal fit is consistent with $z>5.8$ data for no evolution in the hydrogen neutral fraction. 
 However, W07 pointed out that the lognormal fit does not reproduce the data at $5.4<z<5.7$.

 In either case, 
 the large variation in the $\tau$ is the primary reason why determining a more realistic model is difficult.
 Although our results add  $\tau$ measurements at the highest redshift probed by QSO absorption,
 it is clear that we need a statistically significant sample ($>$20) of bright QSOs in every $\Delta z$=0.1 at $z>6$, where $\tau$ quickly increases, to overcome cosmic variation of $\tau$ and more accurately trace the evolution of $\tau$.

\section{Summary}

 Using a deep, medium-resolution Keck/Deimos spectra of a QSO at z=6.4, we measured neutral hydrogen absorption, and thereby, the optical depth.
 Because our QSO is by 2.5 mag fainer than the previously studied QSO SDSSJ1148+5251 at z=6.4, we can investigate absorption much closer to the QSO emission, reaching z=6.365, the highest absorption redshift investigated to date. 
 Our measurement of transmitted flux is consistent with zero at $5.915<z_{abs}^{\alpha}<6.365$ in all \lya, \lyb, and \lyg measurements. We obtained a lower limit of $\tau^{\alpha}>4.8$. 
 These results are consistent with strong evolution of the optical depth at $z>5.7$.

\section*{Acknowledgments}

We thank the anonymous referee for many insightful comments, which significantly improved the paper.
We are grateful to Michael Koss for useful discussion, 
 Hisanori Furusawa and Yutaka Komiyama for providing their Subaru/Suprimecam data of the target.
We thank M. Cooper, J. Newman, and B. Lemaux  for their assistance in using the DEEP2 pipeline.


T.G. acknowledges financial
support from the Japan Society for the Promotion of
Science (JSPS) through JSPS Research Fellowships for Young
Scientists.

This work is supported in part with the research fund for students (2010) of the Department of Astronomical Science, the Graduate University for Advanced Studies, Japan.


%
The authors wish to recognize and acknowledge the very significant cultural role and reverence that the summit of Mauna Kea has always had within the indigenous Hawaiian community.  We are most fortunate to have the opportunity to conduct observations from this sacred mountain.
%
%





\label{lastpage}


\begin{thebibliography}{99}

\bibitem[\protect\citeauthoryear{Becker, Rauch, 
\& Sargent}{2007}]{2007ApJ...662...72B} Becker G.~D., Rauch M., Sargent W.~L.~W., 2007, ApJ, 662, 72 


\bibitem[\protect\citeauthoryear{Becker et al.}{2006}]{2006ApJ...640...69B} 
Becker G.~D., Sargent W.~L.~W., Rauch M., Simcoe R.~A., 2006, ApJ, 640, 69 

\bibitem[\protect\citeauthoryear{Becker et al.}{2011}]{Becker2011} 
Becker G.~D., et al., arXiv:1101.4399


\bibitem[\protect\citeauthoryear{Davis et al.}{2007}]{2007ApJ...660L...1D} 
Davis M., et al., 2007, ApJ, 660, L1 



\bibitem[\protect\citeauthoryear{Faber et al.}{2003}]{2003SPIE.4841.1657F} 
Faber S.~M., et al., 2003, SPIE, 4841, 1657 



\bibitem[\protect\citeauthoryear{Fan et al.}{2006}]{2006AJ....132..117F} 
Fan X., et al., 2006, AJ, 132, 117 


\bibitem[\protect\citeauthoryear{Fan et al.}{2003}]{2003AJ....125.1649F} 
Fan X., et al., 2003, AJ, 125, 1649 




\bibitem[\protect\citeauthoryear{Fan et al.}{2001}]{2001AJ....122.2833F} 
Fan X., et al., 2001, AJ, 122, 2833 


\bibitem[\protect\citeauthoryear{Fan et al.}{2000}]{2000AJ....119..928F} 
Fan X., et al., 2000, AJ, 119, 928 




\bibitem[\protect\citeauthoryear{Goto et al.}{2009}]{2009MNRAS.400..843G} 
Goto T., Utsumi Y., Furusawa H., Miyazaki S., Komiyama Y., 2009, MNRAS, 
400, 843 

\bibitem[\protect\citeauthoryear{Komatsu et 
al.}{2011}]{2011ApJS..192...18K} Komatsu E., et al., 2011, ApJS, 192, 18 


\bibitem[\protect\citeauthoryear{Oh 
\& Furlanetto}{2005}]{2005ApJ...620L...9O} Oh S.~P., Furlanetto S.~R., 2005, ApJ, 620, L9 




\bibitem[\protect\citeauthoryear{Oke 
\& Korycansky}{1982}]{1982ApJ...255...11O} Oke J.~B., Korycansky D.~G., 1982, ApJ, 255, 11 

\bibitem[\protect\citeauthoryear{Pentericci et 
al.}{2002}]{2002AJ....123.2151P} Pentericci L., et al., 2002, AJ, 123, 2151 



\bibitem[\protect\citeauthoryear{Songaila}{2004}]{2004AJ....127.2598S} 
Songaila A., 2004, AJ, 127, 2598 

\bibitem[\protect\citeauthoryear{Songaila 
\& Cowie}{2002}]{2002AJ....123.2183S} Songaila A., Cowie L.~L., 2002, AJ, 123, 2183 


\bibitem[\protect\citeauthoryear{Telfer et al.}{2002}]{2002ApJ...565..773T} 
Telfer R.~C., Zheng W., Kriss G.~A., Davidsen A.~F., 2002, ApJ, 565, 773 



\bibitem[\protect\citeauthoryear{Utsumi et al.}{2010}]{2010ApJ...721.1680U} 
Utsumi Y., Goto T., Kashikawa N., Miyazaki S., Komiyama Y., Furusawa H., 
Overzier R., 2010, ApJ, 721, 1680 

\bibitem[\protect\citeauthoryear{Vanden Berk et 
al.}{2001}]{2001AJ....122..549V} Vanden Berk D.~E., et al., 2001, AJ, 122, 
549 


\bibitem[\protect\citeauthoryear{Willott et 
al.}{2010}]{2010AJ....140..546W} Willott C.~J., et al., 2010, AJ, 140, 546 


\bibitem[\protect\citeauthoryear{Willott et 
al.}{2007}]{2007AJ....134.2435W} Willott C.~J., et al., 2007, AJ, 134, 2435 


\bibitem[\protect\citeauthoryear{Wirth et al.}{2004}]{2004AJ....127.3121W} 
Wirth G.~D., et al., 2004, AJ, 127, 3121 


\end{thebibliography}
\end{document}